\documentstyle[twocolumn,prl,epsfig,aps]{revtex}

\begin{document}
\draft
\title{Inter-isotope determination of ultracold rubidium interactions from
three high-precision experiments}
\author{E.G.M. van Kempen$^1$, S.J.J.M.F. Kokkelmans$^{1*}$, D.J. Heinzen$^2$,
and B.J. Verhaar$^1$}
\address{$^1$Eindhoven University of Technology, P.O. Box 513, 5600 MB
 Eindhoven,
The Netherlands\\
$^2$Department of Physics, University of Texas, Austin, Texas 78712}

\wideabs{

\maketitle

\begin{abstract}
Combining the measured binding energies of four of the most weakly bound
 rovibrational
levels of the $^{87}$Rb$_2$ molecule with the results of two other recent
high-precision rubidium experiments, we obtain exceptionally strong constraints
 on the atomic
interaction parameters in a highly model independent analysis. The comparison of
$^{85}$Rb and $^{87}$Rb data, where the two isotopes are related by a mass
 scaling
procedure, plays a crucial role. Using the consistent picture of the
 interactions
that thus arises we are led to predictions for scattering lengths, clock shifts,
Feshbach resonance fields and widths with an unprecedented level of accuracy. To
demonstrate this, we predict two Feshbach resonances in mixed-spin scattering
 channels
at easily accessible magnetic field strengths, which we expect to play a role in
the damping of coherent spin oscillations.
\end{abstract}


}

After the first realization of Bose-Einstein condensation (BEC) in a
dilute ultracold gas of rubidium atoms\cite{anderson}, experiments with
the two isotopes $^{87}$Rb and $^{85}$Rb further lead to an amazingly
rich variety of BEC phenomena, ranging from the controlled collapse of a
condensate with tunable attractive interactions\cite{roberts1} to
the realization of an atomic matter wave on a microchip\cite{reichel}.
Because of the large number of groups that have started doing experiments with
these atomic species and the growing complexity and
subtlety of the planned experiments, there is a clear need for a more precise
knowledge of the interactions between ultracold rubidium atoms in the
electronic ground state, since these determine most of the properties of
the condensate. For instance, despite a widespread interest, until now to
our knowledge no experimental group has been able to locate the
predicted\cite{vogels} magnetic-field induced Feshbach resonances that can
be used to tune the interactions between ultracold $^{87}$Rb atoms.
Being able to switch on or off these interactions at will by a mere change
of magnetic field may well be one of the main assets of matter waves compared
to light waves in the new matter wave devices. In an atomic interferometry
device, in particular, a nonlinear interaction between interfering waves
may be introduced or eliminated by changing a field applied at
the intersection point.

In this Letter, combining the results of three very recent high-precision
observations, we come close to a complete and model-independent specification
of the interaction properties of ultracold rubidium atoms. The fact that two
isotopes $^{85}$Rb and $^{87}$Rb are involved in the measurements makes the
 constraints
exceptionally strong and also increases the predictive power: the interaction
properties of any other fermionic or bosonic isotope with mass number 82, 83,
84, or 86 are now known with about the same precision. Using mass scaling to
relate the different isotopes we are able for the first time to deduce for each
 of the
isotopes the exact numbers of bound Rb$_{2}$ states with total spin $S$ = 0
(singlet) and 1 (triplet). As an illustration of the predictive power we predict

two Feshbach resonances in mixed-spin scattering channels for $^{87}$Rb
at easily accessible fields that could lead to new time dependent phenomena
in coherent spin oscillations and spin waves. There are numerous effects, such
as spinor condensate energy differences, which are proportional to differences
of scattering lengths. Because these differences are unusually small in Rb,
the potentials must be very accurate to calculate them to reasonable accuracy.

The first of the three high-precision experiments is the recent measurement
of four of the highest bound rovibrational levels of the $^{87}$Rb$_2$
molecule with 10 kHz precision\cite{freeland}. The second experiment is
the improved characterization\cite{roberts} of the elastic scattering near
a Feshbach resonance in $^{85}$Rb, leading to a more precise determination of
 the
resonance field $B_0$ = 154.9(4) G and the nearby field
strength $B'_0 \equiv B_0 + \Delta$ = 165.85(5) G, where the
scattering length goes through zero ($\Delta$ is the (elastic) resonance
width). The third experimental ingredient going
into our analysis is the measurement\cite{seto} of 12148 transition
frequencies between X$^1\Sigma^+_g$ vibrational levels of the ($^{85}$Rb)$_2$,
($^{87}$Rb)$_2$, and $^{85}$Rb$^{87}$Rb molecules, leading to a highly
accurate singlet Rb + Rb potential\cite{pureC6}. Moreover,
within the accuracy of this experiment a comparison of levels for the three
studied isotopomers shows no sign of Born-Oppenheimer break-down effects,
i.e., the observed levels agree with a simple radial Schr\"{o}dinger equation
containing a common singlet potential $V_S(r)$ and the reduced atomic mass.

This set of extremely precise measurements calls for a very careful
construction of the interatomic total spin $S$ = 0 and 1 potentials,
depending on the interatomic separation $r$. We
combine the singlet potential of Ref.~\cite{seto} with a long-range part
equal to the difference $V_{disp} - V_{exch}$ of a dispersion term and an
exchange term, starting at a variable radius $r_S$ between 21 and 23.5 $a_0$
(1 $a_0$ = 0.529{\AA}).
The part $V_{disp}(r)$ includes $C_6, C_8, C_{10}$ terms and retardation,
while $V_{exch}(r)$ is given by the analytic form
$\frac{1}{2} J r^{7/2\alpha - 1} \exp(-2\alpha r)$ derived by Smirnov and
Chibisov\cite{smirnov}, in which $\frac{1}{2} \alpha^2$ is the ionization
potential of the Rb atom in atomic units (au).

The triplet potential is subject to a larger uncertainty. For its
short range part an {\em ab initio} potential is usually taken. To get rid of
 this
model dependence, we use the accumulated phase method\cite{moerdijk}:
the 'history' of the atom-atom motion is summarized by a boundary
condition at an interatomic distance $r_0$, in the form of the phase
$\phi_T(E,l)$ of the oscillating triplet radial wave function $\psi$
depending on energy $E$ and angular momentum $l$. Specifying $\phi_T(E,l)$
is equivalent to giving the logarithmic derivative $\psi'/\psi$ at $r=r_0$.
In all of our previous work we neglected the singlet-triplet mixing by
the hyperfine interaction $V_{hf}$ of the nuclear and electronic spins
in the range $r < r_0$, in order to deal with pure singlet and triplet
radial waves until the boundary. Here, however, we introduce a new variant that
allows us to choose a larger $r_0$ than would otherwise be possible: we
include the adiabatic mixing by $V_{hf}$ in the two-atom spin states but
still neglect its influence on the radial wave functions to avoid dependence
on the history other than via the pure triplet phase. Model calculations
show that in this form the scattering calculations have the required
accuracy for $r_0$ values up to 16 $a_0$. The experimental data for either
ultracold or weakly bound atoms that we analyze
comprise a small $E$ and $l$ range near $E = l = 0$. In this range a first
order Taylor expansion $\phi_T(E,l) = \phi_T^0 + E \phi_T^E + l(l+1) \phi_T^l$
is adequate, which reduces the information contained in $V_T(r)$ for $r < r_0$
 to
three phase parameters only. In principle, these would be needed for both
the $^{85}$Rb and  $^{87}$Rb systems. However, since we
expect Born-Oppenheimer breakdown effects to be negligible also for the
triplet channel in the distance range $r < r_0$, we use mass scaling to
express $\phi_T^0, \phi_T^E, \phi_T^l$ for $^{85}$Rb in terms of the three
phase parameters for $^{87}$Rb. Beyond $r_0$ we construct $V_T(r)$ from
$V_S(r)$ by adding $2V_{exch}(r)$.

Applying this method we carry out a full quantum scattering calculation
for a set of eight experimentally measured quantities. This set
consists of five quantities for $^{87}$Rb and three for
$^{85}$Rb. The $^{87}$Rb data are the four bound state energies
and the ratio of scattering lengths $a_{1-1}/a_{21} = 1.062(12)$ for atomic
scattering in condensates of $^{87}$Rb atoms in the hyperfine states $(f,m_f) =
(1,-1)$ and $(2,1)$ \cite{matthews}.
For $^{85}$Rb we include the Feshbach resonance fields $B_0$ and
$B'_0$, as well as the energy 0.7(1) mK of the g-wave shape resonance
observed in the scattering of a pair of cold atoms in the total spin $S = 1$
state\cite{boesten}.

With a least-squares search routine we determine optimal
values for the parameters $C_6, C_8, J, \phi_T^0(^{87}$Rb), $\phi_T^E(^{87}$Rb),
$\phi_T^l(^{87}$Rb). $C_{10}$ is kept fixed at the value calculated by Marinescu
{\em et al.}~\cite{marinescu}, but the effect of $\pm 10\%$ variations around
 this
value and an estimated upper bound for the influence of higher dispersion
terms are included in the final error bars.
Column A summarizes the main results of the calculations. We
find a value for $C_6$ in agreement with the theoretical value 4691(23) obtained
by Derevianko {\em et al.}~\cite{derevianko}. The $C_8$ value agrees with that
calculated by Marinescu {\em et al.}\cite{marinescu}. To our knowledge this
is the first experimental determination of $C_8$. Our analysis also yields the
first experimental value of the strength of $V_{exch}$ for cold atoms. The
coefficient $J$ agrees with the most recent theoretical value in
 Ref.~\cite{hadinger}.
Table \ref{table1} also gives the values of the pure singlet and triplet
 scattering
lengths for both $^{85}$Rb and $^{87}$Rb, following from
$C_6, C_8, J, \phi_T^0(^{87}$Rb), as well as the fractional
vibrational quantum numbers at dissociation $v_D$ and the numbers
of bound states $n_b$. The reduced minimum $\chi^2$ value is 0.5.

The foregoing makes clear that a major step forward has been made possible by
the new experiments, two of which make use of a Bose-Einstein condensate. This
 is
a firm basis for making a variety of interesting predictions. As a first example
we predict the $^{87}$Rb $f = 1$ spinor condensate to be ferromagnetic, i.e., it
is favorable for two $f = 1$ atoms to have their spins parallel, because the
 mean
field interaction is more repulsive for total $F = 0$ than for $F = 2$: the
 calculated
scattering lengths are $a_{F=2} = +100.4(1) a_0$ and $a_{F=0} = +101.8(2) a_0$.
In a recent preprint Klausen {\em et al.}\cite{klausen} independently come to
 this
conclusion of a ferromagnetic spinor condensate by calculating the scattering
lengths for several assumed numbers of triplet bound states.

We are also able to predict collisional frequency shifts in an
$^{87}$Rb fountain clock for arbitrary choices of partial
densities of atomic hyperfine states. Table~\ref{table2} compares
our calculated fractional frequency shifts normalized to total
atom density $n$ for two recent experiments\cite{fertig,sortais}.
We find good agreement with the three measured shifts.

For various applications there is widespread interest for
predictions of magnetic field values at which Feshbach resonances are
to be expected in the scattering of two $^{87}$Rb atoms in the
$(f,m_f) = (1,+1)$ state. With our interaction parameters we expect them
at the four resonance field values $B_0$ given in Table~\ref{table3} together
with the widths $\Delta$. The $B_0$ values are to be compared with the values
383, 643, 850, and 1018 G predicted in 1997 \cite{vogels}. It is interesting
that the broadest resonance at 1004 G shows a doublet structure\cite{vankempen}.

\begin{figure}[h]
\begin{center}\
  \epsfxsize=80mm \epsfbox{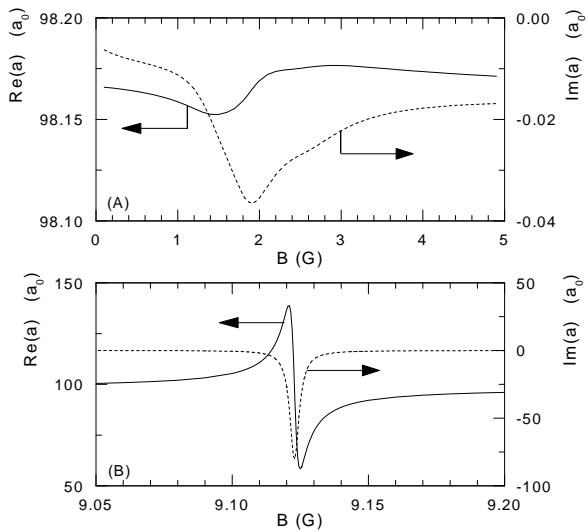}
\end{center}
\caption{(A) Real (solid line) and imaginary (dashed line) parts
of the scattering length $a(B)$ in s-wave $^{87}$Rb (2,+1)+(1,-1)
mixed spin scattering channel, showing the presence of a Feshbach
resonance at 1.9G. Imaginary part is proportional to summed rate
coefficient $G$ for decay into all open channels. (B) Same for
Feshbach resonance in (2,-1)+(1,+1) channel at 9.1G.} \label{fig1}
\end{figure}

Figure \ref{fig1} shows Feshbach resonances that we predict to occur in the
mixed spin channels (2,+1)+(1,-1) and (2,-1)+(1,+1) at easily
accessible field values of 1.9 and 9.1G, respectively. The graphs
show the predicted field-dependent scattering lengths $a(B)$,
which are complex functions due to the presence of exothermal
inelastic decay channels. The generalized analytic expression for
the field dependence in this case is\cite{vankempen}:
\begin{equation}
a(B) = a_\infty \left( 1-e^{2i\phi_R} \frac{\Delta_{el}}%
{B-B_0+\frac{1}{2}i\Delta_{inel}} \right),
\end{equation}
with $\Delta_{el}$ and $\Delta_{inel}$ the (in)elastic resonance
widths and $\phi_R$ a resonance phase constant, arising due to
inelasticity. Note that the real part of the scattering length
does not go through infinity. It turns out that the 1.9G resonance
is an $l$ = 2 resonance, which couples via the spin-spin
interaction $V_{ss}$ to the s-wave incident channel. Actually,
this resonance is the `hyperfine analog state'\cite{vankempen} in
the (2,+1)+(1,-1) scattering channel of the d-wave shape resonance
of $^{87}$Rb occurring in the spin-stretched (2,+2)+(2,+2) spin
channel\cite{boesten}, i.e., a state with essentially the same
spatial dependence and differing only in its hyperfine spin
structure\cite{cornell}. It is located at a comparable low energy
above threshold. In a similar way the $l$ = 0 resonance at 9.1G is
the hyperfine analog state of two of the $l$ = 0 bound states
observed\cite{freeland} at roughly 25MHz below threshold in the
(2,+2)+(2,+2) and (1,-1)+(1,-1) channels, belonging to the same
rotational band as the d-wave shape resonance. They might play a
role in the damping of coherent spin oscillations of the type
which are being observed in experiments at JILA\cite{cornell1}.

Until now we assumed the expression for $V_{disp}$ to be valid for
interatomic distances larger than $r_S$. We now extend $V_{disp}$
with $C_{11}$ and $C_{12}$ terms and assume it to be valid also
between 18 $a_0$ and $r_S$. This leads us to a more ambitious
approach that allows us to determine $C_{10}$ and $C_{11}$ as two
more free parameters in the least squares search: we take into
account the additional constraint arising from the equality
$V_S(r) = V_{disp}(r) - V_{exch}(r)$ by imposing this equality at
five $r$ points as additional 'experimental data' with a  standard
deviation of 0.5\%. We thus effectively include the bound states
of Ref.~\cite{seto} with outer turning points in the range
considered. In the search we take $C_{12}$ equal to the
theoretical value $11.9 \times 10^{9}$ of Ref.~\cite{patil}. In
column B of Table~\ref{table1} the resulting optimal parameter
values are given together with error bars based on a 25 \%
uncertainty in $C_{12}$. We find a value for $C_{10}$ differing
from the theoretical value $7.665 \times 10^7$ au \cite{marinescu}
by only 1.8\%. While the above attractive $C_n$ terms with even
$n$ arise from the interatomic multipole-multipole interaction in
second order, a $C_{11}$ term is expected\cite{dalgarno} as a
repulsive third order dispersion term arising from the mutual
dipole excitation and deexcitation of the atoms with an
intermediate quadrupole transition between excited states in each
of the atoms. Note that the ratio $C_{11}/C_{12} = -0.072$ is
comparable to the rigorous value -0.028 for H atoms\cite{bukta}
and the {\em ab initio} ratio -0.041 for Cs
atoms\cite{weickenmeier}. The remaining residue of the fit,
concentrated at the smallest radii in the radial interval may well
be due to the summed contributions of further (attractive and
repulsive) dispersion terms beyond the $C_{12}$ contribution plus
correction terms to the Smirnov-Chibisov exchange expression. Note
that the values of the lower dispersion coefficient are dominated
by the close-to-threshold measurements, whereas the higher ones
are determined primarily by the Seto potential in the middle range
$r_0 < r < r_S$. We expect that experiment will prove the value of
this more ambitious approach.

For completeness we point out that a weak contribution to the total atom-atom
force is still missing in the above picture: the interatomic spin-spin
interaction $V_{ss}$. One component of $V_{ss}$ is the well-known magnetic
 dipole
interaction between the valence electron spins of the interacting atoms. An
additional contribution, which arises from the electronic spin-orbit coupling
as a second-order effect, has been experimentally determined for the first time
for rubidium atoms by Freeland {\em et al.}\cite{freeland}. Calculation shows
that $V_{ss}$ has a negligible influence on the previous analysis.

In summary, combining the results of three recent high-precision experiments
we have come close to a complete and model independent specification
of the interaction properties of cold rubidium atoms. We have determined
the van der Waals coefficients $C_6$, $C_8$, $C_{10}$, $C_{11}$, and the
strength $J$ of the exchange interaction. We have thus reached a consistent
picture of the interactions, with which it is possible to predict essentially
all parameters needed for a complete description of a rubidium Bose-Einstein
condensate or thermal gas of any isotope in an arbitrary spin state.
New experimental data, in particular on the Feshbach resonances, will
undoubtedly be helpful to confirm the above consistent picture and to further
narrow down the error limits. We believe that our approach sets an example for
similar experimental and theoretical work for other (combinations of) atomic
species. From a theoretical point of view, it is fascinating that it is possible
to achieve a level of precision for the interaction properties approaching that
 for
hydrogen atoms, based on a combination of experimental results and a
sound framework of collision physics. Additional details and their relevance for
future experiments will be the subject of a future publication\cite{vankempen}.

We gratefully acknowledge the support of the work at Texas by the R.A. Welch
Foundation, the US National Science Foundation, and the NASA Microgravity
Research Division. The work at Eindhoven is part of the research program of the
Stichting FOM, which is financially supported by NWO.

\begin{table} {\hspace{2.7cm} Results of analysis}
\begin{center}
\begin{tabular}[t]{clll}
 Quantity & A & B \\
     \hline
$C_6/10^3$        & 4.703(9) & 4.698(4)  \\
$C_8/10^5$        & 5.79(49) & 6.09(7)    \\
$C_{10}/10^7$     & 7.665(Ref.~\cite{marinescu})  & 7.80(6)   \\
$C_{11}/10^9$     & - & -0.86(17)   \\
$C_{12}/10^9$     & - & 11.9(Ref.~\cite{patil})   \\
$J.10^2$          & 0.45(6) & 0.42(2)  \\
$a_T(^{87}$Rb)    & +98.98(4)& +98.99(2)  \\
$a_S(^{87}$Rb)    & +90.4(2) & +90.0(2)  \\
$a_T(^{85}$Rb)    & -388(3) & -387(1)  \\
$a_S(^{85}$Rb)    & +2795$^{+420}_{-290}$     & +2400$^{+370}_{-150}$  \\
$v_{DT},n_{bT}(^{87}$Rb) & 40.4215(3),\ 41 & 40.4214(2),\ 41  \\
$v_{DS},n_{bS}(^{87}$Rb) & 124.455(1),\ 125 & 124.456(1),\ 125  \\
$v_{DT},n_{bT}(^{85}$Rb) & 39.9471(2),\ 40 & 39.9470(1),\ 40  \\
$v_{DS},n_{bS}(^{85}$Rb) & 123.009(1),\ 124 & 123.011(1),\ 124  \\
\end{tabular}
\end{center}
\caption{Interaction parameters derived from experiments without (column A)
and including (column B) the requirement $V_S = V_{disp}-V_{exch}$ for
$r_0 < r < r_S$.}
\label{table1}
\end{table}

\begin{table} {\hspace{2cm} Predicted clock frequency shifts}
\begin{center}
\begin{tabular}[t]{lll}
 $\left (\frac{1}{n} \frac{\Delta \nu}{\nu} \right )_{exp}$ (10$^{-24}$cm$^3$) &

Ref. & Present theory (10$^{-24}$cm$^3$) \\
     \hline
-56$^{+84}_{-21}$ & \cite{fertig} & -72.5 $\pm$ 3.3 \\
-50(10)$^{+22}_{-34}$ & \cite{sortais} & -32.8 $\pm$ 0.7 \\
-60(16)$^{+29}_{-46}$ & \cite{sortais} & -41.5 $\pm$ 2.9 \\
\end{tabular}
\end{center}
\caption{Predictions of collisional frequency shifts for the $^{87}$Rb fountain
clock, compared to two recent experiments.}
\label{table2}
\end{table}

\begin{table} {\hspace{1cm} Predicted $^{87}$Rb (1,+1) Feshbach resonances}
\begin{center}
\begin{tabular}[t]{cllll}
 $B_0(G)$ & 403(2) & 680(2) & 899(4) & 1004(3) \\
 $\Delta(mG)$ & $<$ 1 & 15 & $<$ 5 & 216 \\
\end{tabular}
\end{center}
\caption{Resonance fields $B_0$ and widths $\Delta$.}
\label{table3}
\end{table}


\begin{references}
\bibitem{pureC6} The analysis in Ref.~\cite{seto} uses a pure
$C_6$ dispersion tail at $r$ values where our analysis shows higher
dispersion terms to be significant. Calculation shows that this affects
the derived singlet potential up to radii of about 23.5 $a_0$ to a
negligible extent. Note that the transition frequencies define relative
potential values only.
\bibitem{cornell} A relation to the d-wave shape resonance was suggested to us
 by
Eric Cornell.
\end{references}
\end{document}